\documentclass[conference]{IEEEtran}
\IEEEoverridecommandlockouts

\usepackage{graphicx}
\usepackage{booktabs}
\usepackage{amsmath}
\usepackage{amssymb}
\usepackage{subcaption}
\usepackage{etoolbox}
\usepackage{multirow}
\usepackage{algorithm}
\usepackage{algorithmic}
\usepackage{pgfplots}
\usepackage{pgfplotstable}
\usepgfplotslibrary{groupplots}

\pgfplotsset{
    discard if not/.style 2 args={
        x filter/.code={
            \edef\tempa{\thisrow{#1}}
            \edef\tempb{#2}
            \ifx\tempa\tempb
            \else
                
            \fi
        }
    }
}
\pgfplotsset{compat=newest}

\DeclareMathOperator*{\argmax}{arg\,max}

\usepackage{hyperref}

\usepackage{tikz}
\usetikzlibrary{decorations.pathmorphing}
\usetikzlibrary{decorations.pathreplacing}
\usetikzlibrary{calc}
\usetikzlibrary{fit}
\usetikzlibrary{matrix}

\definecolor{nicegreen}{RGB}{39,174,96}
\definecolor{niceblue}{RGB}{52,152,219}
\definecolor{niceorange}{RGB}{243,156,18}
\definecolor{nicered}{RGB}{231,76,60}
\definecolor{niceyellow}{RGB}{241,196,15}

\newtheorem{definition}{Definition}

\begin{document}

\title{GPU-Accelerated Optimizer-Aware Evaluation of Submodular Exemplar Clustering}

\author{\IEEEauthorblockN{Philipp-Jan Honysz}
\IEEEauthorblockA{\textit{Artificial Intelligence Unit} \\
\textit{TU Dortmund University}\\
Dortmund, Germany \\
philipp.honysz@tu-dortmund.de}
\and
\IEEEauthorblockN{Sebastian Buschjäger}
\IEEEauthorblockA{\textit{Artificial Intelligence Unit} \\
\textit{TU Dortmund University}\\
Dortmund, Germany \\
sebastian.buschjaeger@tu-dortmund.de}
\and
\IEEEauthorblockN{Katharina Morik}
\IEEEauthorblockA{\textit{Artificial Intelligence Unit} \\
\textit{TU Dortmund University}\\
Dortmund, Germany \\
katharina.morik@tu-dortmund.de}
}

\maketitle

\begin{abstract}
The optimization of submodular functions constitutes a viable way to perform clustering. Strong approximation guarantees and feasible optimization w.r.t. streaming data make this clustering approach favorable. Technically, submodular functions map subsets of data to real values, which indicate how ``representative'' a specific subset is. Optimal sets might then be used to partition the data space and to infer clusters. Exemplar-based clustering is one of the possible submodular functions, but suffers from high computational complexity. However, for practical applications, the particular real-time or wall-clock run-time is decisive.  In this work, we present a novel way to evaluate this particular function on GPUs, which keeps the necessities of optimizers in mind and reduces wall-clock run-time. To discuss our GPU algorithm, we investigated both the impact of different run-time critical problem properties, like data dimensionality and the number of data points in a subset, and the influence of required floating-point precision. In reproducible experiments, our GPU algorithm was able to achieve competitive speedups of up to 72x depending on whether multi-threaded computation on CPUs was used for comparison and the type of floating-point precision required. Half-precision GPU computation led to large speedups of up to 452x compared to single-precision, single-thread CPU computations. 
\end{abstract}

\section{Introduction}
\label{sec:introduction}

Clustering is the most important learning task in unsupervised learning, which aims at identifying meaningful subgroups in data by exploiting dissimilarity or distance between individual observations. A lot of different algorithms have been proposed, that introduce various notions of what clusters are and how they may be discovered in data. Procedures like DBSCAN \cite{Ester/etal/1996a} or OPTICS \cite{Ankerst/etal/1999a} derive clusters directly from data by inspecting the neighborhood of individual points, whereas e.g. $k$-means clustering minimizes a specific loss function by employing an EM-style algorithm. The optimization of submodular functions constitutes another option for identifying clusters in data. From a practical point of view, these functions aim at measuring the ``utility'' or the ``representativity'' of a particular subset of data. Furthermore, they maintain a property of diminishing returns, which during optimization leans towards the identification of rather compact \textit{summarizations} of the observed data. Found representative data points in summaries may then be exploited to derive clusters, serving as their respective cluster ''exemplars''. Clustering using submodular functions might be more favorable than conventional methods, as optimization is possible with strong theoretical guarantees and is also feasible in streaming data settings that require inherently real-time processing.

Submodular functions like the Informative Vector Machine \cite{Lawrence/etal/2002a, Badanidiyuru/etal/2014a} have been proposed, but require the specification of a positive-definite Mercer kernel, which, depending on the concrete data, might be cumbersome. Exemplar-based clustering, as a submodular function, is more flexible as it allows for arbitrary, non-negative dissimilarity functions and enables end-users to include domain knowledge they might have from ''traditional'' clustering. However, this particular function is expensive to compute as the time complexity is $\mathcal{O}(n \cdot k)$ for datasets of size $n$ and subsets (i.e. the set of desired cluster exemplars) of size $k$. This problem gains additional importance as optimizers for submodular functions usually evaluate $l$ sets for their function value when choosing the next optimization step, which leads to a total time complexity of $\mathcal{O}(n \cdot k \cdot l)$. 

In this work, we want to present a novel method to accelerate the evaluation of Exemplar-based clustering. To accomplish this acceleration we employ Graphics Processing Units (GPU), which are wide-spread co-processors and well-established to deal with massively parallelizable tasks. We also discuss, how the necessities of submodular function optimizers might be incorporated into the algorithm design and which hardware pitfalls have to be kept in mind to exploit the computational power of GPUs at best. 

Our contributions are as follows:
\begin{itemize}
    \item We present the first algorithm to evaluate Exemplar-based clustering on GPUs and discuss, how this procedure exploits hardware features like shared memory and coalesced access to minimize runtime. 
    \item We show, how the procedure is capable of running under low-memory conditions, which are common among GPUs.
	\item We conduct a series of experiments to compare different CPU implementations with our GPU algorithm and determine the possible benefits w.r.t. the achievable run-time and speedup.
\end{itemize}

This paper is organized as the following: In section \ref{sec:related_work} we give a short overview of practical applications to submodular functions and of current acceleration strategies for clustering. In section \ref{sec:submodular_functions}, we formally establish submodular functions and the important cardinality-constrained optimization problem. Furthermore, we introduce the Greedy optimizer. In section \ref{sec:exemplar_based_clustering} we will establish the submodular function of Exemplar-based clustering, briefly explain how it measures representativity, discuss, how an implementation for CPUs might look like and which acceleration possibilities are feasible, before introducing our GPU algorithm. In section \ref{sec:experiments} we present our experiments and the achieved results. In section \ref{sec:conclusion} we summarize our work and give an outlook on how this work might be refined.

\section{Related Work}
\label{sec:related_work}

Submodular functions have proven their relevance to practical problems in several use cases: They have already been employed to select features among frequent subgraphs \cite{Borgwardt/etal/2009a}, to empirically validate software \cite{Chen/etal/2017a}, to detect events in activity networks \cite{Rozenshtein/etal/2014a}, and to place sensors effectively \cite{Bellala/etal/2012a, Bryan/etal/2019a}. Submodular functions are also well-known tools to summarize data sets or data streams and select representative observations from them. Their use has been proposed to summarize video data \cite{Gygli/etal/2015a, Mirzasoleiman/etal/2018a}, image data \cite{Tschiatschek/etal/2014a}, and text corpora \cite{Lin/etal/2009a, Lin/Bilmes/2011a, Sipos/etal/2012a}. The optimization of submodular functions is a theoretically well-studied subject \cite{Nemhauser/etal/78a, Nemhauser/Wolsey/78a} with algorithms like ThreeSieves \cite{Buschjaeger/etal/2020b}, SieveStreaming \cite{Badanidiyuru/etal/2014a}, SieveStreaming++ \cite{Kazemi/etal/2019a} and Salsa \cite{NorouziFard/etal/2018a} providing strong guarantees towards the achievable function value and the ability to optimize with streaming data. To the best of our knowledge, the acceleration of Exemplar-based clustering as a submodular function has not been discussed so far. The acceleration of clustering, however, has been (among others) discussed for density-based clustering \cite{Andrade/etal/2013a}, $k$-means clustering \cite{Bhimani/etal/2015a}, $k$-medoids, and self-organizing maps \cite{Kohlhoff/etal/2011a} but not for Exemplar-based clustering. 

\section{Submodular Functions}
\label{sec:submodular_functions}

Submodular functions are set functions, which introduce a property of diminishing returns. To provide an adequate description of this characteristic, the \textit{discrete derivative} is being formalized as follows \cite{Krause/Golovin/2012a}:
\begin{definition}[Discrete derivative]
Let $V$ be a finite set (usually known as the ``ground set''), $f \colon \mathcal{P}(V) \rightarrow \mathbb{R}$, $S \subseteq V$ and $e \in V$. Then, $\Delta_f(e \mid S) = f(S \cup \left\lbrace e \right\rbrace) - f(S)$ represents the \textit{discrete derivative} of $f$ and the set $S$ w.r.t. the observation $e$.
\end{definition}
This definition directly leads to the formalization of submodular functions \cite{Krause/Golovin/2012a}:
\begin{definition}[Submodular function]
A function $f\colon \mathcal{P}(V) \rightarrow \mathbb{R}$ is \textit{submodular}, iff for all $A \subseteq B \subseteq V$ and $e \in V \setminus B$ the following condition holds:
\begin{align}
\Delta_f(e \mid A) \geq \Delta_f(e \mid B)
\end{align}
\end{definition}
This definition does not enforce monotonicity, which, however, is often required for optimization with strong guarantees. Hence, it is established in the following way \cite{Krause/Golovin/2012a}:
\begin{definition}[Monotonic submodular functions]
A submodular function $f$ is \textit{monotone}, iff for all $A \subseteq B \subseteq V$ it holds, that $f(A) \leq f(B)$.
\end{definition}

The problem of submodular function maximization is usually regarded in its cardinality constrained formulation, which we are also discussing here:
\begin{align}
S_k^* = \underset{S \subseteq V, |S| \leq k}{\max} f(S)
\label{eqn:submodular_function_optimization_cardinality_constraint}
\end{align}
Finding the global optimal solution to problem \ref{eqn:submodular_function_optimization_cardinality_constraint} is NP-hard \cite{Krause/Golovin/2012a} as this would require evaluating an exponential number of sets for their function value. 

\begin{algorithm}
   \caption{Greedy optimization}
   \label{alg:greedy}
	\begin{algorithmic}
		\STATE \textbf{Input}: Ground set $V$, $k \in \mathbb{N}$, Monotone submodular function $f$
		\STATE $S_0 \leftarrow \emptyset$
		\FOR{$i \in [1, k]$}
			\STATE $S_i = S_{i-1} \cup \left\{\argmax_{e \in V \setminus S_{i-1}} \Delta_f(e \mid S_{i-1}) \right\}$
		\ENDFOR
		\STATE \textbf{return} $S_k$
	\end{algorithmic}
\end{algorithm}

It is well-known that the optimal function value with a polynominal number of function evaluations is $(1 - e^{-1}) \approx 63,21\%$ \cite{Nemhauser/Wolsey/78a}. Interestingly, this approximation is achieved with a simple Greedy algorithm \cite{Nemhauser/etal/78a}, which is depicted in \ref{alg:greedy}.

\section{Exemplar-based Clustering}
\label{sec:exemplar_based_clustering}

$k$-medoids is a clustering task, that aims at selecting $k$ data points from a data set, which are subsequently used to infer clusters. Exemplar-based clustering is strongly related to $k$-medoids clustering and its accompanying loss function, which is subject to minimization and may be defined as follows \cite{Gomes/Krause/2010a}:
\begin{definition}[$k$-medoids loss]
Let $\mathcal{X}$ be some data space, $V \subseteq \mathcal{X}$ a finite set, $d: \mathcal{X} \times \mathcal{X} \rightarrow \mathbb{R}^+$ some distance function and $S \subseteq V$. The $k$-medoids loss is then defined by
\begin{align}
L(S) = |V|^{-1} \sum_{\vec{v} \in V} \min_{\vec{s} \in S} d(\vec{v}, \vec{s})
\label{eqn:k_medoids_loss}
\end{align}
\end{definition}

One can formulate that problem as a monotone submodular function \cite{Gomes/Krause/2010a}, to profit from optimization in streaming settings and strong guarantees w.r.t. the achievable function value.
\begin{definition}[Exemplar-based clustering]
Let $L: \mathcal{X} \rightarrow \mathbb{R}^+$ be the $k$-medoids loss function. Furthermore, let $\vec{e}_0 \in \mathcal{X}$ be some auxiliary vector, e.g. the all-zero vector $\vec{e}_o = (0,\dots,0)^T$. Then, \textit{exemplar-based clustering} may be defined as a monotone submodular function, as follows:
\begin{align}
f(S) = L(\left\lbrace \vec{e}_0 \right\rbrace) - L(S \cup \left\lbrace \vec{e}_0 \right\rbrace)
\label{eqn:exemplar_based_clustering}
\end{align} 
\end{definition}

Equation \ref{eqn:exemplar_based_clustering} measures the representativity of a selected set $S$ by computing the mean distance of all vectors to their nearest representative: If these distances are very high on average -- which is the case, when data vectors are very dissimilar to their nearest representative -- then, the $k$-medoids loss becomes larger. Exemplar-based clustering now measures the difference between a ``dummy'', very unrepresentative set, which is given by the auxiliary vector $\left\lbrace \vec{e}_0 \right\rbrace$, and between the actual set of representatives $S$. The larger this difference becomes, the more representative $S$ is. Hence, it is also obvious, that the representativity of $S$ is maximal, when $S = V$. Conversely, the representativity is minimal, when $S = \emptyset$, since $f(S) = L(\left\lbrace \vec{e}_0 \right\rbrace) - L(\left\lbrace \vec{e}_0 \right\rbrace) = 0$. A beneficial property of Exemplar-based clustering is, that distance functions are not required to compute eq. \ref{eqn:k_medoids_loss} as long as the given function $d$ preserves non-negativity \cite{Badanidiyuru/etal/2014a}. Hence, it is also conceivable, to construct dissimilarity functions from Mercer kernels, if it is beneficial to the problem to solve.

\subsection{Evaluation on CPUs}
\label{sec:exem_eval_cpu}

\begin{algorithm}
   \caption{Exemplar-based clustering (CPU)}
   \label{alg:exemplar_based_clustering_cpu}
	\begin{algorithmic}
		\STATE \textbf{Input}: Datasets $V$ and $S \subseteq V$, Dissimilarity function $d$
		\FUNCTION{$L(V, S)$}
			\STATE $\Sigma \leftarrow \textbf{new} \text{ array } \textbf{of} \text{ size = } |V|$
			\FORALL{$\vec{v}_i \in V$}
				\STATE $t \leftarrow $\texttt{FLT\_MAX}
				\FORALL{$\vec{s} \in S$}
					\STATE $t \leftarrow min(t, d(\vec{s}, \vec{v}_i))$
				\ENDFOR
				\STATE $\Sigma_i \leftarrow t$
			\ENDFOR
			\STATE $\sigma \leftarrow$ \textbf{reduce} $\Sigma$ \textbf{by} sum
			\STATE \textbf{return} $|V|^{-1}\sigma$
		\ENDFUNCTION
		\STATE \textbf{return} $L(V, \left\lbrace \vec{e}_0 \right\rbrace) - L(V, S \cup \left\lbrace \vec{e}_0 \right\rbrace)$
	\end{algorithmic}
\end{algorithm}

To evaluate equation \ref{eqn:exemplar_based_clustering} one can consider a simple procedure as depicted in algorithm \ref{alg:exemplar_based_clustering_cpu}. This na\"ive implementation has a runtime complexity of $\mathcal{O}(|V| \cdot |S|)$. This is especially problematic for large $V$ and $S$, hence reducing runtime in practice is desirable. It is natural to consider index-structures to accelerate heavily distance-dependent algorithms. In fact, algorithm \ref{alg:exemplar_based_clustering_cpu} is a brute-force method, which may be characterized by a central nearest neighbor-query (NN queries). Index structures, like $k$-$d$-trees, require that they are built upon some subset of the data space, which then can be queried for nearest neighbors \cite{Yianilos/1993a}. For equation \ref{eqn:k_medoids_loss} this would require establishing an index on the set $S$, which during optimization changes for \textit{every} function evaluation. Hence, we do not consider the use of index structures.

It is possible to accelerate algorithm \ref{alg:exemplar_based_clustering_cpu} by parallelizing the outer loop, which computes partial sums for every $\vec{v}_i \in V$ and therefore leads to $\mathcal{O}(|V|)$ computational tasks (single set parallelized problem). However it has to be noted, that a large group of optimizers, like the Greedy algorithm, Sieve Streaming \cite{Badanidiyuru/etal/2014a} or Salsa \cite{NorouziFard/etal/2018a}, do not consider a single set $S$ for evaluation but multiple sets $S_\text{multi} = \left\lbrace S_1, \dots, S_l \right\rbrace$ in every optimization step (multiset parallelized problem). For example, consider the Greedy algorithm: In every optimization step the procedure decides, which not yet selected item from the ground set leads to the greatest marginal gain in terms of function value. Given the currently optimal set $S_i$ and a set of candidate items $C = \left\lbrace \vec{c}_1, \dots, \vec{c}_m \right\rbrace$, which might be selected in the current optimization step, this leads to $S_\text{multi} = \left\lbrace S_i \cup \left\lbrace \vec{c}_1 \right\rbrace, \dots, S_i \cup \left\lbrace \vec{c}_m \right\rbrace \right\rbrace$, which needs to be evaluated for their respective function values. While $\mathcal{O}(|V|)$ computational tasks already do not seem to be feasible for contemporary multi-core systems, the multiset parallelized problem imposes $\mathcal{O}(|V| \cdot |S_\text{multi}|)$ tasks, which even more stresses the need for hardware that is better suited to solve that massive number of independent computational problems. This is especially true, since $|C| \approx |V|$ during Greedy optimization. 

\subsection{Evaluation on GPUs}

GPUs are a very popular choice to accelerate massively parallel tasks due to its many-core architecture, its strong focus on SIMD-like jobs, and its general availability to the wide public. Hence, we choose the GPU to accelerate the multiset parallelized problem of Exemplar-based clustering and discuss an appropriate implementation. Although different models for general purpose GPU (GPGPU)-programming exist, we will focus on the CUDA framework throughout this work.

While it is possible to organize threads on CPUs in virtually arbitrary ways (e.g. by employing job pools, producer-consumer systems etc.), for GPUs we have to map the specific job into a grid-block structure. Every \textit{grid} may have up to three dimensions, which in turn contains \textit{blocks}. Likewise, every block may also have up to three dimensions and consists of no more than 1024 threads. It is important to note, that threads are issued in \textit{warps}, which represent a group of usually 32 threads that are launched together at the same address and process the same instruction. Every thread from a single block has access to comparably small \textit{shared memory}, which in contrast to \textit{global memory} represents on-chip memory. Hence, accesses to shared memory are relatively fast.

We establish a formal framework to discuss the mapping into the grid-block structure, as follows: Let $C = (D_g, D_b)$ be a specific \textit{kernel configuration} to solve a particular problem, with $D_g = (g_x, g_y, g_z)$ and $D_b = (b_x, b_y, b_z)$ the dimensioning of the grid and the block respectively. Every thread knows, which position $P_t = (t_x^*, t_y^*, t_z^*)$ it acquires in a specific block $P_b = (b_x^*, b_y^*, b_z^*)$. This information is essential, since every individual thread derives from it which task it has to conduct.

\subsubsection{Algorithm}

For our GPU implementation we consider equation \ref{eqn:exemplar_based_clustering} again: It can be seen, that the expression $L(\left\lbrace \vec{e}_0 \right\rbrace)$ is independent of the given set $S$ to evaluate. Since that sub-expression is $\mathcal{O}(|V|)$ in terms of time complexity, we do not consider any acceleration and compute that term conventionally, which makes the resulting value available to \textit{all} subsequent computations. 

We will focus on the expression $L(S \cup \left\lbrace \vec{e}_0 \right\rbrace)$ from now on: To ensure, that we can place as many independent tasks on the GPU as possible, we consider a decomposition of the aforementioned sub-expression, as follows:
\begin{align}
		\label{eqn:loss_decomposed}
        L_{\vec{v}_i} (S) = |V|^{-1} \min_{\vec{s} \in S} d(\vec{v}_i, \vec{s}) \\
		\label{eqn:loss_decomposed_summarized}
        L(S) = L_{\vec{v}_1} (S) + \cdots + L_{\vec{v}_n} (S)
\end{align}
Using that decomposition and the set of sets to evaluate $S_\text{multi}$ we can construct a work matrix $\mathbf{W}$, which constitutes the foundation to design the grid-block structure from:
\begin{align}
        \mathbf{W} = \begin{pmatrix}
        L_{\vec{v}_1}(S_1) & L_{\vec{v}_2}(S_1) & \cdots & L_{\vec{v}_n}(S_1) \\ 
        L_{\vec{v}_1}(S_2) & L_{\vec{v}_2}(S_2) & \cdots & L_{\vec{v}_n}(S_2) \\ 
        \vdots & \vdots  & \ddots &  \vdots \\ 
        L_{\vec{v}_1}(S_l) & L_{\vec{v}_2}(S_l) & \cdots &  L_{\vec{v}_n}(S_l)
        \end{pmatrix}
		\label{eq:work_matrix}
\end{align}
In this work matrix, the $i$-th row corresponds to the $i$-th set from $S_\text{multi}$, whereas the $j$-th column belongs to the $j$-th element from the ground set $V$. It can be seen, that a reduction of equation \ref{eq:work_matrix} by sum in column direction leads back to equation \ref{eqn:loss_decomposed_summarized}, which is the result we are ultimately looking for. We design our procedure so that every GPU thread is assigned to a single cell from $\mathbf{W}$. 

As algorithm \ref{alg:exemplar_based_clustering_cpu} already suggests, many memory accesses are made w.r.t. the elements from $V$. Therefore, this should be performed as fast as possible. Hence, we load vectors $\vec{v}_i \in V$ from global memory into shared memory, that serves as a low latency, user-managed cache. Because shared memory is block exclusive, we strive to put as many threads $L_{\vec{v}_i}$ into one block as possible. From the perspective of equation \ref{eq:work_matrix}, that means that we want to maximize the block dimensions in row direction (first matrix dimension, ``$y$ direction''). To accomplish this, we have to provide an appropriate block dimensioning $D_b$, which considers both the restrictions of the programming model (max. 1024 threads per block) and the number of bytes $\beta$ we may allocate per block from shared memory. Let $\gamma$ be the number of bytes every $\vec{v}_i \in V$ requires to be stored, then we can determine $D_b = (b_x, b_y, b_z = 1)$ to be as follows:
\begin{align*}
b_x = \min \left\lbrace \left\lfloor \frac{1024}{b_Y} \right\rfloor, \left\lfloor \frac{\beta}{\gamma} \right\rfloor \right\rbrace \;
b_y = \min \left\lbrace 1024, |S_{\text{multi}}| \right\rbrace
\end{align*}
It can be seen that $b_y$ represents the maximization we have discussed before: For a single vector from $V$, we  create as many threads in the $y$ direction as we have sets in $S_\text{multi}$ but not more than 1024 threads. Conversely, $b_x$ is constructed by instantiating as many threads as we have still left after considering $b_y$. Additionally, if the block is growing in ``$x$ direction'' (second matrix dimension, cf. equation \ref{eq:work_matrix}), we are taking new ground vectors into account. Hence, we also have to take care, that we do not exceed the per-block shared memory size limitation $\beta$. 

To provide a complete kernel configuration $C = (D_g, D_b)$, we now have to specify a grid dimensioning $D_g$. From a given block dimensioning $D_b$ we construct $D_g = (g_x, g_y, g_z = 1)$ such that every cell from $\mathbf{W}$ is being computed:
\begin{align}
g_x = \left\lceil \frac{|V|}{b_X} \right\rceil \; g_y = \left\lceil \frac{|S_{\text{multi}}|}{b_Y} \right\rceil
\end{align}
The routine will computationally ensure, that no threads will be executed which do not match to any cell of the work matrix $\mathbf{W}$.

\begin{algorithm}
        \caption{Exemplar-based clustering (GPU)}
        \begin{algorithmic}
				\REQUIRE{dissimilarity function $d$}
                \STATE{$i \leftarrow b_x \cdot b_x^* + t_x^*$}
                \STATE{$j \leftarrow b_y \cdot b_y^* + t_y^*$}
                \IF{$t_y^* = 0$}
                        \STATE{\textbf{load} $\vec{v}_i \in V$ \textbf{from} global memory \textbf{into} shared memory}
                \ENDIF
                \STATE{\textbf{synchronize} threads \textbf{in} block}
                \STATE{$d_\text{min} \leftarrow $ \texttt{FP\_MAX}}
                \FORALL{$\vec{s}_i \in S_j$}
                        \STATE{$d_\text{min} \leftarrow \min(d_\text{min}, d(\vec{s}_i, \vec{v}_i))$}
                \ENDFOR
                \STATE{$W_{j, i} = \frac{d_\text{min}}{|V|}$}
        \end{algorithmic}
        \label{alg:exemplar_based_clustering_gpu}
\end{algorithm}

As we have established the kernel configuration, we will now focus on the concrete kernel, i.e. the computational routine to calculate every cell from $\mathbf{W}$. The kernel is depicted in algorithm \ref{alg:exemplar_based_clustering_gpu}: First, every thread determines to which cell of $\mathbf{W}$ it belongs to by calculating the $i$ and $j$ index, i.e. the column and the row of $\mathbf{W}$ respectively. Then, every first thread in $y$ direction loads the accompanying vector from global into shared memory. During that operation, every other thread in the same block waits by employing a barrier synchronization. The following lines compute equation \ref{eqn:loss_decomposed}, whereby the final result is ultimately written back to the work matrix, which resides in global memory. The resulting work matrix is then reduced in a row-wise fashion on the GPU by calculating $\mathbf{W} \cdot \vec{1}$, which delivers the final result for every $S \in S_\text{multi}$.

\subsubsection{Memory Layout}

So far we are assuming that our computational payload (i.e. $V$ and $S_\text{multi}$) \textit{somehow} has been copied and stored in global memory. To ensure efficiency, we have to copy the payload in as few transactions as possible. Furthermore, we have to present the data in an advantageous way to the GPU. Let us now discuss this in more detail.

CPU and GPU memory do not differ much when it comes to accessing techniques. Every application has access to some allocated address space, in which a one-dimensional address indicates, where some information should be read from or written to. Hence, a vector can easily be stored by successively writing its contents to memory. However, matrices represent \textit{complex} data structures since they require two-dimensional addressing across rows and columns. Therefore, we need to map matrices to ``ordinary'' vectors. Let us call this process \textit{vectorization}. The usual approaches to vectorization are row- or column-wise storage, where a specific matrix is either successively stored by its rows or its columns, respectively. For our algorithm, we store the $\mathbf{V}$ matrix in a column-wise fashion and copy it in a single memory transaction to the GPU. Since the ground matrix never changes between different function evaluations it is copied to the GPU's global memory on algorithm initialization. We do not consider further optimization, because no further memory accesses on the GPU, beyond loading vectors from global to shared memory, are conducted, as discussed above.

\begin{figure}
    \centering
    \begin{subfigure}{0.495\textwidth}
            \resizebox{\textwidth}{!}{%
                    \begin{tikzpicture}
                    \foreach \x in {0,...,0}{
                            \path[draw] (2 * \x, 0) rectangle ++(2, 0.5);
                            \foreach \y in {0,...,3}
                            \path[draw, fill=nicered!50] (2 * \x + \y * 0.5 , 0) rectangle ++(0.5, 0.5);
                    }
                    \foreach \x in {1,...,3}{
                            \path[draw] (2 * \x, 0) rectangle ++(2, 0.5);
                            \foreach \y in {0,...,3}
                            \path[draw, fill=nicegreen!50] (2 * \x + \y * 0.5 , 0) rectangle ++(0.5, 0.5);
                    }
                    
                    \node[] at (0, -0.25) {0};
                    \node[] at (2, -0.25) {32};
                    \node[] at (4, -0.25) {64};
                    \node[] at (6, -0.25) {128};
                    
                    \foreach \x in {0,...,3}
                    \draw [->](0.25 + \x * 0.5, 0.75) -- (0.25 + \x * 0.5, 0.5);
                    \end{tikzpicture}
            }
            \subcaption{Sequential access.}
            \label{fig:CUDA_SequentialAccess}
    \end{subfigure}
    \begin{subfigure}{0.495\textwidth}
            \resizebox{\textwidth}{!}{%
                \begin{tikzpicture}
                \foreach \x in {0,...,3}{
                    \path[draw] (2 * \x, 0) rectangle ++(2, 0.5);
                    \foreach \y in {1,...,3}
                            \path[draw, fill=nicegreen!50] (2 * \x + \y * 0.5 , 0) rectangle ++(0.5, 0.5);
                            \foreach \y in {0,...,0}
                            \path[draw, fill=nicered!50] (2 * \x + \y * 0.5 , 0) rectangle ++(0.5, 0.5);
            }
    
            \node[] at (0, -0.25) {0};
            \node[] at (2, -0.25) {32};
            \node[] at (4, -0.25) {64};
            \node[] at (6, -0.25) {128};
            
            \foreach \x in {0,...,3}
                    \draw [->](0.25 + \x * 2, 0.75) -- (0.25 + \x * 2, 0.5);
                    \end{tikzpicture}
            }
            \subcaption{Strided access.}
            \label{fig:CUDA_StridedAccess}
    \end{subfigure}
    \caption{Example GPU memory area consisting of four segments of 32 bytes each. \textbf{(a)} Four 8 byte words are requested from a single memory segment of the global memory, resulting in exactly one memory transaction. \textbf{(b)} Four 8-byte words are requested from four memory segments of the global memory, resulting in four memory transactions.}
	\label{fig:coalescing}
\end{figure}
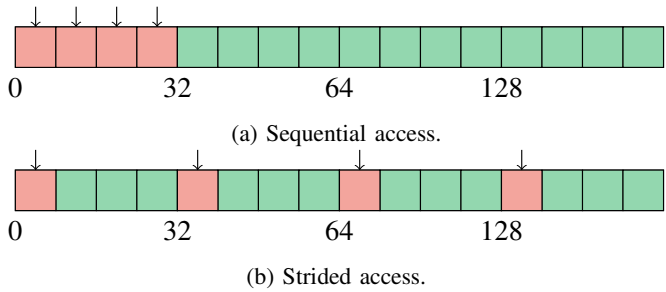

This is different when it comes to the set of evaluation matrices $S_\text{multi}=\left\lbrace \mathbf{S}_1, \dots, \mathbf{S}_l \right\rbrace$ since they are too large to be cached in shared memory. Therefore, access by global memory is mandatory and has to be conducted as efficiently as possible. In CUDA, global memory is divided into segments of fixed size, which usually is 32 bytes. Therefore a segment allows for up to four 64-bit floating-point numbers (FP64) or up to eight 32-bit floating-point numbers (FP32). If threads of a single warp access a particular memory segment, then these memory accesses become \textit{coalesced} into a single memory transaction, which is beneficial to runtime and data throughput. Conversely, if threads of a single warp access different memory segments then more memory transactions are needed to access the same data. An example of these considerations is given in figure \ref{fig:coalescing}.

\begin{figure}
	\centering
	\resizebox{\columnwidth}{!}{%
	\begin{tikzpicture}[
			matrixstyle/.style={matrix of nodes, nodes in empty cells, column sep=-\pgflinewidth, row sep=-\pgflinewidth, nodes={inner sep=0mm,outer sep=0pt, minimum size=5mm, text height=\ht\strutbox,text depth=\dp\strutbox, draw}}, 
			brace/.style={decoration={brace}, decorate},
   			position label/.style={above = 2pt, text height = 2ex, text depth = 1ex}
			]
		
		\matrix (S1) at (0, 30mm) [matrixstyle, ampersand replacement=\&]{
			|[fill=nicegreen!50]| $\cdot$ \& |[fill=nicegreen!50]| $\cdot$ \& |[fill=nicegreen!50]| $\cdot$ \& |[fill=nicegreen!50]| $\cdot$ \\
			|[fill=nicegreen!50]| $\cdot$ \& |[fill=nicegreen!50]| $\cdot$ \& |[fill=nicegreen!50]| $\cdot$ \& |[fill=nicegreen!50]| $\cdot$ \\
		};
	
		\matrix (S2) at (25mm, 30mm) [matrixstyle, ampersand replacement=\&]{
			|[fill=niceblue!50]| $\cdot$ \& |[fill=niceblue!50]| $\cdot$ \& |[fill=niceblue!50]|$\cdot$ \\
			|[fill=niceblue!50]| $\cdot$ \& |[fill=niceblue!50]| $\cdot$ \& |[fill=niceblue!50]|$\cdot$ \\
		};
	
		\matrix (S3) at (52.5mm, 30mm) [matrixstyle, ampersand replacement=\&]{
			|[fill=niceyellow!50]| $\cdot$ \& |[fill=niceyellow!50]| $\cdot$ \& |[fill=niceyellow!50]| $\cdot$ \& |[fill=niceyellow!50]| $\cdot$ \& |[fill=niceyellow!50]| $\cdot$ \\
			|[fill=niceyellow!50]| $\cdot$ \& |[fill=niceyellow!50]| $\cdot$ \& |[fill=niceyellow!50]| $\cdot$ \& |[fill=niceyellow!50]| $\cdot$ \& |[fill=niceyellow!50]| $\cdot$ \\
		};

		\node[] (S3DBrTop) at ($(S3-1-5.north east)+(0.3,-0.2)$) {};
		\node[] (S3DBrBottom) at ($(S3-2-5.south east)+(0.3,0.1)$) {};
		
		\node[] (S1BrLeft) at ($(S1-1-1.north west)+(0.0,0.325)$) {};
		\node[] (S1BrRight) at ($(S1-1-4.north east)+(0.0,0.325)$) {};
		
		\node[] (S2BrLeft) at ($(S2-1-1.north west)+(0.0,0.325)$) {};
		\node[] (S2BrRight) at ($(S2-1-3.north east)+(0.0,0.325)$) {};

		\node[] (S3BrLeft) at ($(S3-1-1.north west)+(0.0,0.325)$) {};
		\node[] (S3BrRight) at ($(S3-1-5.north east)+(0.0,0.325)$) {};

		\draw [brace] (S3DBrTop.north) -- node [pos=0.5, xshift=3.5mm] {$d$} (S3DBrBottom.south);
		\draw [brace] (S1BrLeft.south) -- node [position label, pos=0.5] {$|S_1|$} (S1BrRight.south);
		\draw [brace] (S2BrLeft.south) -- node [position label, pos=0.5] {$|S_2|$} (S2BrRight.south);
		\draw [brace] (S3BrLeft.south) -- node [position label, pos=0.5] {$|S_3|$} (S3BrRight.south);
		
		\matrix (SMat) at (27.5mm, 5mm) [matrixstyle, ampersand replacement=\&]{
			|[fill=nicegreen!50]| $\cdot$ \&|[fill=niceblue!50]| $\cdot$ \& |[fill=niceyellow!50]| $\cdot$ \& 
			|[fill=nicegreen!50]| $\cdot$ \&|[fill=niceblue!50]| $\cdot$ \& |[fill=niceyellow!50]| $\cdot$ \& 
			|[fill=nicegreen!50]| $\cdot$ \&|[fill=niceblue!50]| $\cdot$ \& |[fill=niceyellow!50]| $\cdot$ \&
			|[fill=nicegreen!50]| $\cdot$ \&                     $\cdot$ \& |[fill=niceyellow!50]| $\cdot$ \&
                                  $\cdot$ \&                     $\cdot$ \& |[fill=niceyellow!50]| $\cdot$ \\
			|[fill=nicegreen!50]| $\cdot$ \&|[fill=niceblue!50]| $\cdot$ \& |[fill=niceyellow!50]| $\cdot$ \& 
			|[fill=nicegreen!50]| $\cdot$ \&|[fill=niceblue!50]| $\cdot$ \& |[fill=niceyellow!50]| $\cdot$ \& 
			|[fill=nicegreen!50]| $\cdot$ \&|[fill=niceblue!50]| $\cdot$ \& |[fill=niceyellow!50]| $\cdot$ \&
			|[fill=nicegreen!50]| $\cdot$ \&                     $\cdot$ \& |[fill=niceyellow!50]| $\cdot$ \&
                      $\cdot$ \&                     $\cdot$ \& |[fill=niceyellow!50]| $\cdot$ \\
		};

		\node[xshift=2mm, yshift=-5mm, position label] (SMatLabel) at (SMat.east) {$\mathbf{S}$};
		
		\path[draw=black!10, ->] (S1-2-2.south) to [out = 270, in = 90, looseness = 0.5] (SMat-1-4.north);
		\path[draw=black!10, ->] (S1-2-3.south) to [out = 270, in = 90, looseness = 0.5] (SMat-1-7.north);
		\path[draw=black!10, ->] (S1-2-4.south) to [out = 270, in = 90, looseness = 0.5] (SMat-1-10.north);
		
		\path[draw=black!10, ->] (S2-2-2.south) to [out = 270, in = 90, looseness = 0.5] (SMat-1-5.north);
		\path[draw=black!10, ->] (S2-2-3.south) to [out = 270, in = 90, looseness = 0.5] (SMat-1-8.north);
	
		\path[draw=black!10, ->] (S3-2-2.south) to [out = 270, in = 90, looseness = 0.5] (SMat-1-6.north);
		\path[draw=black!10, ->] (S3-2-3.south) to [out = 270, in = 90, looseness = 0.5] (SMat-1-9.north);
		\path[draw=black!10, ->] (S3-2-4.south) to [out = 270, in = 90, looseness = 0.5] (SMat-1-12.north);
		\path[draw=black!10, ->] (S3-2-5.south) to [out = 270, in = 90, looseness = 0.5] (SMat-1-15.north);
		\draw[->] (S1-2-1.south) -- (SMat-1-1.north);
		\draw[->] (S2-2-1.south) to [out = 270, in = 90, looseness = 1] (SMat-1-2.north);
		\draw[->] (S3-2-1.south) to [out = 270, in = 90, looseness = 0.5] (SMat-1-3.north);
		
		\matrix (SMatVec) at (27.6mm, -17.5mm) [matrixstyle, ampersand replacement=\&]{
			|[fill=nicegreen!50]| $\cdot$ \&|[fill=niceblue!50]| $\cdot$ \& |[fill=niceyellow!50]| $\cdot$ \& 
			|[fill=nicegreen!50]| $\cdot$ \&|[fill=niceblue!50]| $\cdot$ \& |[fill=niceyellow!50]| $\cdot$ \& 
			|[fill=nicegreen!50]| $\cdot$ \&|[fill=niceblue!50]| $\cdot$ \& |[fill=niceyellow!50]| $\cdot$ \&
			|[fill=nicegreen!50]| $\cdot$ \&                     $\cdot$ \& |[fill=niceyellow!50]| $\cdot$ \&
                                  $\cdot$ \&                     $\cdot$ \& |[fill=niceyellow!50]| $\cdot$ \&
			|[fill=nicegreen!50]| $\cdot$ \&|[fill=niceblue!50]| $\cdot$ \& |[fill=niceyellow!50]| $\cdot$ \& 
			|[fill=nicegreen!50]| $\cdot$ \&|[fill=niceblue!50]| $\cdot$ \& |[fill=niceyellow!50]| $\cdot$ \& 
			|[fill=nicegreen!50]| $\cdot$ \&|[fill=niceblue!50]| $\cdot$ \& |[fill=niceyellow!50]| $\cdot$ \&
			|[fill=nicegreen!50]| $\cdot$ \&                     $\cdot$ \& |[fill=niceyellow!50]| $\cdot$ \&
                      $\cdot$ \&                     $\cdot$ \& |[fill=niceyellow!50]| $\cdot$ \\
		};

		\path[draw=black, ->] (SMat-2-7.south) -- node [text width=2.5cm,midway,right] {Vectorization} ($(SMat-2-7.south)-(0.0,1.5)$);
	
		\node[] (T1a) at ($(SMatVec-1-1.south west)-(-0.25,0.75)$) {$t_1$};
		\node[] (T2a) at ($(SMatVec-1-2.south west)-(-0.25,0.75)$) {$t_2$};
		\node[] (T3a) at ($(SMatVec-1-3.south west)-(-0.25,0.75)$) {$t_3$};
		\node[] (T1b) at ($(SMatVec-1-4.south west)-(-0.25,0.75)$) {$t_1$};
		\node[] (T2b) at ($(SMatVec-1-5.south west)-(-0.25,0.75)$) {$t_2$};
		\node[] (T3b) at ($(SMatVec-1-6.south west)-(-0.25,0.75)$) {$t_3$};
		\node[] (Tdots) at ($(SMatVec-1-7.south west)-(-0.375,0.75)$) {$\cdots$};
		
		\draw[->] (T1a.north)  -- (SMatVec-1-1.south);
		\draw[->] (T2a.north)  -- (SMatVec-1-2.south);
		\draw[->] (T3a.north)  -- (SMatVec-1-3.south);
		\draw[->] (T1b.north)  -- (SMatVec-1-4.south);
		\draw[->] (T2b.north)  -- (SMatVec-1-5.south);
		\draw[->] (T3b.north)  -- (SMatVec-1-6.south);
	\end{tikzpicture}
	}
	\caption{Three evaluation matrices $\mathbf{S}_1$, $\mathbf{S}_2$ and $\mathbf{S}_3$ with four, three and five elements respectively and a dimensionality of $d=2$ are being processed into a single matrix $\mathbf{S}$. Subsequently, the resulting matrix is being \textit{vectorized} in a row-wise fashion. The threads $t_1$, $t_2$ and $t_3$ are assigned to the evaluation matrices $\mathbf{S}_1$, $\mathbf{S}_2$ and $\mathbf{S}_3$ and successively access the information of vectorized matrix, which leads to coalesced access and to as few memory transactions as possible.}
	\label{fig:ExemGPU_BuildSummaryMatrix}
\end{figure}
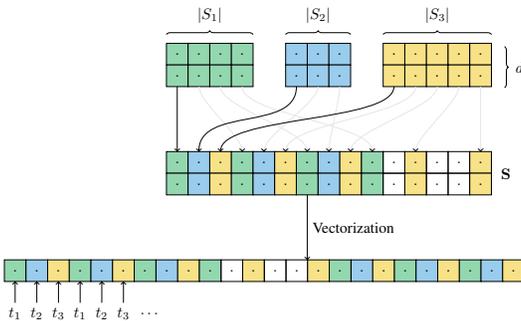

In addition to exploiting mentioned locality effects, we also want to copy the evaluation matrices in as few requests as possible. Copying computational payload is a relatively expensive operation and the bandwidth of the PCI-E connection (which usually serves as the link between CPU and GPU) is only exhausted when sufficiently large data is transmitted. Hence, we propose the following vectorization routine, which not only is aware of locality effects but also seeks to exhaust the bandwidth of the interconnect between CPU and GPU. 

To discuss our vectorization routine, we again take a look at algorithm \ref{alg:exemplar_based_clustering_gpu}. Especially the calculation of the dissimilarity value between two vectors $\vec{s}_i$ and $\vec{v}_i$ is of interest since this is the portion of the code, where data accesses happen. Most dissimilarity functions require to loop over the elements of the vectors to compare. Since $\vec{v}_i$ is already loaded into shared memory, we focus on the various memory accesses to $\vec{s}_i$ from different GPU kernel threads. As we are seeking to optimize accesses of threads of a particular warp, we can assume that every thread processes the same instruction. Let loading $\vec{s}_i[k] \in S_j$ from memory be now the instruction of interest, whereas $k$ indexes a particular dimension. From work matrix $\mathbf{W}$ we can assume, that $i$ and $k$ are equal within threads of the same warp (cf. equation \ref{eq:work_matrix}). To exploit coalesced memory access, we have to optimize loading vectors from $S_j$, whereas $j$ varies for different threads of the same warp and the same block. Hence, the data of the different sets $S_j$ for which $k$ and $i$ are equal must be stored sequentially in memory. From a technical point of view, this can be achieved by choosing an evaluation set $S_j$ in round robin-fashion and selecting the next, not yet processed vector from that set. The selected vector is then written to an evaluation set matrix $\mathbf{S}$, which contains the data of all evaluation sets. If all vectors from a chosen set have been written to the matrix, the entry in $\mathbf{S}$ simply remains empty. Therefore, it has to be noted, that the procedure is not absolutely space-efficient, as blank fields yield unused but allocated memory. Nevertheless, this is convenient for the algorithm, as no variable evaluation set sizes need to be considered, which simplifies addressing the needed data in GPU memory. However, this also means that this effect does not have to be considered if sets do not differ in size. This is the case for the Greedy algorithm (cf. section \ref{sec:exem_eval_cpu}) but not for Sieve-based procedures, like ThreeSieves \cite{Buschjaeger/etal/2020b}.

\subsubsection{Chunking}
\label{sec:gpu_algo_chunking}

While for CPU computations we can usually assume that enough memory is available to solve a specific problem, this assumption may not hold for GPUs due to lack of memory expandability. Hence, in those cases it might be needed to break the presented problem instance into smaller sub-problems or \textit{chunks}. To accomplish chunking, we assess how much memory $\varphi$ is left on the GPU device first. Then, we calculate how much memory $\mu_s$ is needed to solve a problem consisting of a single evaluation set first. This includes the needed space to store $\mathbf{S}$, $\mathbf{W}$ and its metadata but \textit{not} $\mathbf{V}$ since it has been pre-loaded on algorithm initialization and is already considered in $\varphi$. Using $\mu_s$ we are able to determine, how many evaluation sets would fit into a single chunk given the currently free GPU memory $\varphi$, by calculating $n_\text{chunk-size} = \lfloor \varphi \mu_s^{-1} \rfloor$. The number of total chunks is then given by $n_\text{chunks} = \lceil l n_\text{chunk-size}^{-1} \rceil$ with $l$ the number of evaluation sets. The algorithm then splits the set of evaluation sets into $n_\text{chunks}$ of $n_\text{chunk-size}$ each, processes them independently and merges the results after processing all chunks. Chunking fails, when $n_\text{chunk-size}$ equals zero, which means, that there is no memory left to even process a single evaluation set. This might be the case, when $\mathbf{V}$ is already very large, which suggests either the use of lower floating-point precision (reducing the required memory to solve problem instances) or better suited hardware with larger memory.

\section{Experiments}
\label{sec:experiments}

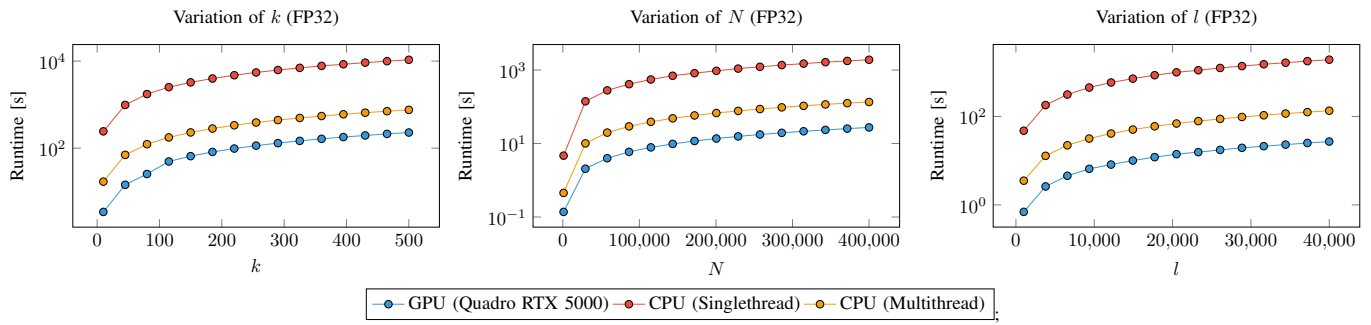
\begin{figure*}
	\centering
	\resizebox{\textwidth}{!}{%
		\begin{tikzpicture}
			\begin{groupplot}[group style={group name=myplot, group size=3 by 3, horizontal sep=1.75cm, vertical sep=1.7cm}]
				\def\myPlots{}
				\pgfplotsforeachungrouped \y in {FP32}{
					\pgfplotsforeachungrouped \x in {k, N, l}{
						\eappto\myPlots{
							\noexpand\nextgroupplot[xlabel=$\x$, ylabel={Runtime [s]}, title=Variation of $\x$\;(\y), 
							legend style={at={($(0, 0)+(1cm,1cm)$)},legend columns=4,fill=none,draw=black,anchor=center,align=center}, legend to name=fred,
							scaled x ticks=false, xticklabel style={
								/pgf/number format/fixed,
								/pgf/number format/precision=5
							},
							scaled y ticks=false, yticklabel style={
								/pgf/number format/fixed,
								/pgf/number format/precision=5
							}, y post scale=0.6, ymode=log]

							\noexpand\addplot[color=niceblue, mark=*, mark options={draw=black}] table[x=\x, y=RUNTIME_QUADRO, col sep=comma] {results/\x-\y-speedup.csv};
							\noexpand\addlegendentry{GPU (Quadro RTX 5000)}

							\noexpand\addplot[color=nicered, mark=*, mark options={draw=black}] table[x=\x, y=RUNTIME_ST, col sep=comma] {results/\x-\y-speedup.csv};
							\noexpand\addlegendentry{CPU (Singlethread)}

							\noexpand\addplot[color=niceorange, mark=*, mark options={draw=black}] table[x=\x, y=RUNTIME_MT, col sep=comma] {results/\x-\y-speedup.csv};
							\noexpand\addlegendentry{CPU (Multithread)}
						}
					}
				}
				\myPlots
			\end{groupplot}

		\node[below] at (current bounding box.south) {\pgfplotslegendfromname{fred};};
		\end{tikzpicture}
	}
	\caption{Runtimes of our GPU algorithm and of corresponding single- and multi-threaded CPU algorithms w.r.t. the variation of $k$, $N$ and $l$. The lower, the better.}
	\label{fig:runtimes_perfloss_variation}
\end{figure*}

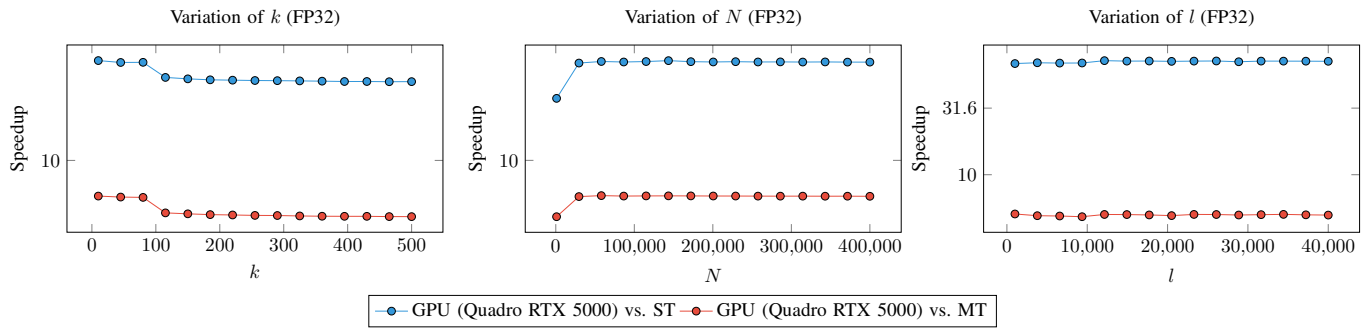
\begin{figure*}
	\centering
	\resizebox{\textwidth}{!}{%
		\begin{tikzpicture}
			\begin{groupplot}[group style={group name=myplot, group size=3 by 2, horizontal sep=1.5cm, vertical sep=1.7cm}]
				\def\myPlots{}
				\pgfplotsforeachungrouped \y in {FP32}{
					\pgfplotsforeachungrouped \x in {k, N, l}{
						\eappto\myPlots{
							\noexpand\nextgroupplot[ymode=log, log ticks with fixed point, xlabel=$\x$, ylabel=Speedup, title=Variation of $\x$\;(\y), 
							legend style={at={($(0, 0)+(1cm,1cm)$)},legend columns=4,fill=none,draw=black,anchor=center,align=center}, legend to name=speeduplegend,
							scaled x ticks=false, xticklabel style={
								/pgf/number format/fixed,
								/pgf/number format/precision=5
							},
							scaled y ticks=false, yticklabel style={
								/pgf/number format/fixed,
								/pgf/number format/precision=5
							}, y post scale=0.6]

							\noexpand\addplot[color=niceblue, mark=*, mark options={draw=black}] table[x=\x, y=Quadro_vs_ST, col sep=comma] {results/\x-\y-speedup.csv};
							\noexpand\addlegendentry{GPU (Quadro RTX 5000) vs. ST}

							\noexpand\addplot[color=nicered, mark=*, mark options={draw=black}] table[x=\x, y=Quadro_vs_MT, col sep=comma] {results/\x-\y-speedup.csv};
							\noexpand\addlegendentry{GPU (Quadro RTX 5000) vs. MT}
						}
					}
				}
				\myPlots
			\end{groupplot}

		\node[below] at (current bounding box.south) {\pgfplotslegendfromname{speeduplegend}};
		\end{tikzpicture}
	}
	\caption{Speedup of our GPU algorithm compared to single- and multi-thread CPU implementations (''ST'' and ''MT''), considering single precision (FP32) floating-point arithmetic.The higher, the better.}
	\label{fig:speedup_variation}
\end{figure*}

In this section, we experimentally evaluate our GPU algorithm and compare it to a CPU implementation, whereby two different variants will be considered: A single-threaded version of algorithm \ref{alg:exemplar_based_clustering_cpu}, which does not consider any parallelization at all, and a multi-threaded version, which runs the mentioned algorithm on different sets in parallel. Both CPU procedures make use of a SIMD strategy, to accomplish the sum reduction described in algorithm \ref{alg:exemplar_based_clustering_cpu}, whereby OpenMP is utilized to achieve that. The evaluation focuses on two research questions:
\begin{enumerate}
\item What is the impact of the number of vectors $N$ in the ground set $V$, the number of subsets $l$, and the number of vectors $k$ in every set $S \in S_\text{multi}$ on the real execution time? From the worst case complexity, we know that these properties bound the run-time. Now, let us see for varying numbers, what the actual impact is.
\item What is the impact of using half-precision (FP16) or single-precision (FP32) on wall-clock runtime?
\end{enumerate}
To discuss these questions adequately, we measure the wall-clock time it took to solve a particular problem, which consists of evaluating some ground set $V$ and a set of subsets $S_\text{multi}$. Every problem is randomly generated, whereby the data generation is \textit{not} part of the measured run-time. All CPU experiments have been conducted on an Intel Xeon W-2155, whereby we considered both 10 physical cores and 10 additional logical cores (from simultaneous multi-threading). For conducting the GPU experiments we chose to compare the performance of the algorithms by employing a workstation-grade NVIDIA Quadro RTX 5000. Furthermore, the squared Euclidean distance $d = \|\vec{x} - \vec{y}\|^2_2$ will be used as a dissimilarity measure between different observations for all our experiments. The reference implementation of the GPU and CPU routines in C++ and Python, which we used to obtain the reported results, is publicly available\footnote{\url{https://github.com/philippjh/exemcl}}.

\subsection{Variation of run-time-critical properties}
\label{sec:experiments_variation_runtime}
\begin{table}
\centering
\begin{tabular}{lllrrr}
\toprule
  &      &    &     min &    mean &     max \\
\midrule
\multirow{4}{*}{$N$} & \multirow{2}{*}{FP16} & ST &    8.47 &  391.31 &  435.96 \\
  &      & MT &    0.82 &   27.71 &   30.50 \\
\cline{2-6}
  & \multirow{2}{*}{FP32} & ST &   33.98 &   67.44 &   71.53 \\
  &      & MT &    3.29 &    4.83 &    4.99 \\
\cline{1-6}
\cline{2-6}
\multirow{4}{*}{$l$} & \multirow{2}{*}{FP16} & ST &  273.94 &  414.53 &  452.47 \\
  &      & MT &   20.34 &   29.23 &   31.69 \\
\cline{2-6}
  & \multirow{2}{*}{FP32} & ST &   68.33 &   70.60 &   71.86 \\
  &      & MT &    4.84 &    4.98 &    5.07 \\
\cline{1-6}
\cline{2-6}
\multirow{4}{*}{$k$} & \multirow{2}{*}{FP16} & ST &   59.64 &  223.89 &  424.11 \\
  &      & MT &    4.21 &   15.81 &   29.91 \\
\cline{2-6}
  & \multirow{2}{*}{FP32} & ST &   46.88 &   52.31 &   71.01 \\
  &      & MT &    3.32 &    3.70 &    4.97 \\
\bottomrule
\end{tabular}

\caption{Achieved minimal, mean and maximal speedup of our GPU algorithm in 15 runs compared to a single-threaded (ST) and multi-threaded (MT) CPU implementation. FP16-GPU speedups were computed from comparison with FP32-CPU wall-clock run-times. As described in section \ref{sec:experiments_variation_runtime}, we report the speedup for different experimental routines, where either $N$, $l$ or $k$ has been varied.}
\label{tab:speedup}
\end{table}

The runtime of the different algorithms is influenced by three factors: The number of observations in the ground set ($N$), the number of subsets to evaluate ($l$) and the number of observations in each of the $l$ subsets ($k$). For this discussion, we initially choose these parameters to be $N = 50000, l = 5000, k = 10$. Now, we consider three series of experiments, in which each of these parameters is varied while every other parameter is left at their initial value. We vary each parameter by choosing 15 uniformly spaced values from a pre-defined interval. Let this interval for $N$ be $[1000, 400000]$, for $l$ be $[1000, 40000]$ and for $k$ be $[10, 500]$. These intervals are chosen, such that no chunking is possible. The dimensionality of every created observation is fixed to $100$. Furthermore, we only consider FP32 computation for this series of experiments since contemporary GPUs suffer in terms of runtime from requiring double precision floating point arithmetics. The results are depicted in figure \ref{fig:runtimes_perfloss_variation} and \ref{fig:speedup_variation} and table \ref{tab:speedup}.

From the depicted graphs and table \ref{tab:speedup} we can derive, that our GPU algorithm is the fastest method to compute Exemplar-based clustering when single-precision arithmetic is used. The achievable speedup compared to the single-thread implementation ranges between 34x in the worst-case scenario and 72x in the best-case scenario. Comparing the GPU algorithm to the multi-threaded CPU implementation shows speedups ranging from 3.29x to 5.07x.

The observed speedups are rather constant for variations of $N$ and $l$, while they are decreasing for growing $k$ (cf. figure \ref{fig:speedup_variation}). An explanation for this behavior could be, that the run-time suffers from loops becoming increasingly larger as more vectors in every evaluation set have to be considered (cf. algorithm \ref{alg:exemplar_based_clustering_gpu}). It is well-known, that GPUs in particular suffer from iterative elements in computational kernels. However, since in submodular optimization we usually seek compact summarizations and only a few cluster exemplars in data, this behavior should not be a problem in practice given that $k$ is comparably small. Beyond that, we could observe outlying speedup behavior for $N = 1000$, which delivered inferior run-times when compared to larger $N$. This might indicate, that the problem size is too small to exhaust the full performance of the tested GPUs. Furthermore, we can observe, that increasing the problem size w.r.t. $k$, $N$ or $l$ leads to quasi-linear increase in run-time, which is what we expect given the theoretical computational complexity of the underlying problem (cf. section \ref{sec:exemplar_based_clustering}).

\subsection{Half-precision run-time benefits}

Second, we want to inspect the impact on run-time, if half-precision arithmetic (FP16) is chosen instead of single-precision arithmetic. Unfortunately, contemporary CPUs do not provide native support for FP16 computations, hence we compare GPU-FP16 results to the CPU-FP32 run-times and take a look at the achievable speedups. For this series of experiments, we consider the same variations of $N$, $l$ and $k$ as in section \ref{sec:experiments_variation_runtime}.

From table \ref{tab:speedup} it can be seen, that FP16 computation yielded speedups of up to 452x, when a single-thread CPU-FP32 algorithm is used for comparison. Compared to the CPU multi-thread implementation, we were able to achieve speedups of up to 32x. In single-thread scenarios, mean speedups for variations of $N$ and $l$ range from 391x to 415x, for variations of $k$ the observable mean speedup equals 224x. Given that the multi-thread CPU algorithm is used, mean speedups for variations of $N$ and $l$ were ranging from 67x to 71x, while the average speedup for variations of $k$ equaled approx. 16x. As already discussed in section \ref{sec:experiments_variation_runtime}, the (comparably) lower mean speedup for variations of $k$ result from the increasing iterative parts of the kernel that are not parallelized and thus slow down computation on GPUs, considerably.

\section{Conclusion}
\label{sec:conclusion}

In this paper, we presented a novel GPU algorithm to evaluate the submodular function of Exemplar-based clustering, which keeps the necessities of submodular optimizers in mind. 
Exemplar-based clustering offers a fast inspection of datasets through representative subsets. However, its complexity of $\mathcal{O}(n \dot k)$ could become prohibitive regarding the wall-clock run-time. 
We briefly discussed the current state of practical applications of submodular functions and GPU accelerations of clustering. We formally established submodular functions and introduced the Greedy optimizer with its approximation guarantees. We then established Exemplar clustering and the accompanying submodular function by discussing $k$-medoids loss. We explained, how this function measures representativity, which is the key to understand which notion of ''cluster'' is employed here. 

We claim that real-time applications demand results on time behavior that is measured in physical entities like seconds. The worst-case complexity determines the influential parameters. Systematic experiments result in wall-clock time for execution and allow to investigate the speedup of one implementation over another in varied settings of the parameters.  

We introduced our novel GPU algorithm, for which we engineered a work matrix first and discussed, how hardware features like shared memory might be used here. We especially took care of choosing an appropriate memory layout and established a way to deal with scarce GPU memory by chunking a given problem into smaller sub-problems. Experiments have shown, that our algorithm succeeds in remarkable speedups of up to 72x using a workstation-grade Quadro RTX 5000 GPU, while assuming FP32 computations and comparing to an adequate single-thread CPU implementation. Given, that a multi-threaded CPU algorithm with 20 worker threads represents the baseline, speedups ranging from 3.29x to 5.07x were possible. Moreover, we also reviewed the usage of half-precision FP arithmetic and revealed great speedups ranging from 16x to 415x compared to the FP32 CPU implementation, depending on whether single- or multi-thread computation was considered and which run-time-critical property was subject to variation. Overall, the memory-aware GPU implementations results in considerable speedups. 

These were especially noticeable when only a single worker thread was able to run on the CPU or when half-precision computations has been considered. 

For future work it might be interesting to see, which impact different floating point precision requirements have towards the found clustering in order to determine, whether FP16 problem solving is viable in real-world scenarios.

\section*{Acknowledgments}

Part of the work on this paper has been supported by Deutsche Forschungsgemeinschaft (DFG) within the Collaborative Research Center SFB 876 "Providing Information by Resource-Constrained Analysis", project A1, \url{http://sfb876.tu-dortmund.de} and by the German Competence Center for Machine Learning Rhine Ruhr (ML2R, \url{https://www.ml2r.de/}, 01IS18038A), funded by the German Federal Ministry for Education and Research.

\bibliography{IEEEabrv, paper}
\bibliographystyle{IEEEtran}

\end{document}